\documentclass[aps,prd,preprint,superscriptaddress,preprintnumbers,nofootinbib]{revtex4-1}

\usepackage{amsmath}
\usepackage{amsfonts} 
\usepackage{amssymb}
\usepackage{graphicx}
\usepackage{hyperref}
\usepackage{tensor}
\usepackage{siunitx}
\usepackage{physics}
\newcommand{\pa}{\partial}
\usepackage{appendix}
\newcommand{\be}{\begin{equation}}
\newcommand{\ee}{\end{equation}}

\usepackage{geometry}
\newcommand{\D}{\text{d}}
\newcommand{\beq}{\begin{equation}}  \newcommand{\eeq}{\end{equation}}
\newcommand{\beqn}{\begin{eqnarray}}
 \newcommand{\eeqn}{\end{eqnarray}}
\newcommand{\bal}{\begin{aligned}}   \newcommand{\eal}{\end{aligned}}
\newcommand{\bea}{\begin{eqnarray}}  \newcommand{\eea}{\end{eqnarray}}
\def\d{\mathrm{d}}

\hypersetup{colorlinks=true,
			urlcolor=purple,
			citecolor=blue,
			linkcolor=magenta,}

\begin{document}

\title{\Large  The Swampland at Large Number of Space-Time Dimensions}
\preprint{LMU-ASC 40/20}
\preprint{MPP-2020-192}
\preprint{DESY 20-185}

{~}

\author{\vspace{0.5cm}
\large Quentin~Bonnefoy\vspace{0.5cm}}
 \affiliation{DESY, Notkestra{\ss}e 85, 22607 Hamburg, Germany\\[1.5ex]}
\author{\large Luca~Ciambelli}
 \affiliation{Universit\' e Libre de Bruxelles and International Solvay Institutes, ULB-Campus Plaine CP231, B-1050 Brussels, Belgium\\[1.5ex]}
\author{\large Dieter~L\"ust}
 \affiliation{Arnold-Sommerfeld-Center for Theoretical Physics, Ludwig-Maximilians-Universit\"at, 80333 M\"unchen, Germany\\[1.5ex]}
\affiliation{Max-Planck-Institut f\"ur Physik (Werner-Heisenberg-Institut),
             F\"ohringer Ring 6,
             80805, M\"unchen, Germany\\[1.5ex]}
\author{\large Severin~L\"ust}
 \affiliation{Jefferson Physical Laboratory, Harvard University, Cambridge, MA 02138, USA}

\begin{abstract}
\vspace{1.0cm}
\noindent
We  discuss some aspects of swampland constraints - 
especially the 
swampland distance conjecture - in a large number of space-time dimensions $D$.
We analyze Kaluza-Klein (KK) states at large $D$ and find that some KK spectra possess an interesting dependence on $D$.
On the basis of these observations we propose a new large dimension  conjecture.
We apply it to KK states of compactifications to anti-de Sitter backgrounds where it predicts an upper bound on the dimension of space-time as a function of its characteristic radius.
We also apply our conjecture to black hole spacetimes, whose entropies have a $D$-dependence very similar to that of the KK spectrum.
\end{abstract}

\maketitle

\section{Introduction}

The so-called swampland program \cite{Vafa:2005ui,Ooguri:2006in,Palti:2019pca} is
addressing the important question about what kind of effective field theories, albeit being apparently fully consistent from the quantum field theory point of view,
cannot be consistently embedded into quantum gravity. One  aspect of the swampland idea   is  that  directions in  field space which are associated to large field values are rejected in quantum gravity 
in the sense that they are accompanied by an infinite  tower of almost massless particles. 

As it is  well-known for many years, string theory as consistent theory of quantum gravity can only exist in a certain critical space-time dimension $D_c$, namely $D_c=10$ for the superstring and the
heterotic strings, $D_c=26$ for the bosonic string and $D_c=11$ for M-theory. There are also certain attempts for non-critical string theories with $D\neq D_c$, however 
their full quantum consistency is still unclear.
The idea of this paper  is to consider the limit of large number of space-time dimensions, $D\rightarrow\infty$, in the context of the swampland distance conjecture  \cite{Ooguri:2006in}.
The large-$D$ limit  and a corresponding $1/D$
expansion of general relativity was already considered before in a series of interesting papers, starting with the work of \cite{Emparan:2013moa} (for a recent review see \cite{Emparan:2020inr}).
Here we  want to present a first attempt to get   a (upper) bound on $D$ via swampland considerations, i.e., we like to address the question:
is large-$D$ quantum gravity in the swampland?

There are no theorems, or even conjectures, classifying quantum gravity theories in terms of space-time dimensions that would settle this question. We do not try to follow this route either. Instead, we study the consistency of known swampland conjectures when they are extended to large space-time dimensions. In particular, we utilize the swampland distance conjecture  \cite{Ooguri:2006in}, 
which is about large distances $\Delta$ in field space of quantum gravity, such as string theory, and about related towers of massless
states that appear at large distances. As a generalization of this idea, 
we propose to determine the distance between space-times of different space-time dimensions. This amounts to enlarge the parameter space of space-time geometries to include, besides the geo\-metric distance parameters (like the compactification radius, the AdS radius or the black hole mass), also the number of dimensions $D$ of space-time as an additional swampland distance parameter. This extension is motivated by the non-trivial behaviour of known towers when they are considered in the large-$D$ limit, which cannot be captured by usual distances on field spaces.
Concretely, we want 
to derive or characterize  new distance functions  $\Delta(D)$ as a function of $D$, as they might provide some possible constraints on the number of space-time dimensions. 

Given that we lack a dynamical theory for the dimension of spacetime, we cannot build a distance for it in the usual way. Instead, we investigate bounds on $D$ from consistency requirements on the towers of states predicted by the SDC, most notably the KK states. We propose to build a distance out of the typical mass of the towers and study its $D$-dependence, using it as a proxy for $\Delta(D)$.

The paper is organized as follows. 
In the next section, we analyze the $D$-dependence of several swampland conjectures.   Then we will propose a {\sl Large Dimension Conjecture (LDC)} that restricts the number of space-time dimensions to be smaller or equal to some critical value that depends on the typical space-time size , or more generally on the value of the moduli.
After that, we focus on towers of KK modes that appear in effective field theories (EFTs) of anti-de Sitter space (AdS), 
and
 we will concretely derive the distance functions $\Delta(D)$  for 
 $AdS_d\times S^{d'}$ backgrounds. We furthermore compute the distance function $\Delta(D)$ for black holes.
 At the end of the paper we will eventually argue a possible new D-duality symmetry between large and small dimension.

\section{Swampland conjectures at large number of dimensions}

The swampland program aims at formulating conditions that must hold for any EFT of a quantum gravity theory. It defines a subset of the seemingly consistent EFTs, 
strictly smaller than the full set,
 but large enough so that the notion of a landscape remains. Within this subset, one may find certain large-$D$ theories. Our goal in this paper is to study the intersection between the landscape and the set of large-$D$ gravity EFTs.

Since classifying all possible quantum theories of gravity is not a viable option, even for backgrounds of fixed dimensions, swampland conjectures have been originally derived by extrapolating properties of known string vacua, or by relying on semi-classical black hole physics. As it is well-known, we cannot use the former at large $D$. Nevertheless, we can choose to take all swampland conjectures at face values, and see how they behave at large $D$. If they become extremely stringent, we would take it as a hint that large-$D$ theories are in the swampland. If on the other hand they stop being constraining, as seems to be the case, the conclusion is less clear. One could conclude that there are simply less theories in the swampland at large $D$. One could also extrapolate the known information that the landscape is much smaller than the set of seemingly consistent EFTs, and state that it should hold at large $D$ as well. Then, we would deduce that either we need more conjectures to restrict the large $D$ landscape within the set of large $D$ EFTs, or that the large-$D$ limit was not allowed in the first place and that large-$D$ theories are in the swampland.

Below, we discuss swampland conjectures which have a well-defined large-$D$ behaviour.

\subsection{The species bound at large D}

Let us start with conjectures that can be motivated independently of string theory and which feature a dimensional dependence. The first one is the species bound, namely the fact that EFTs must have a cutoff $\Lambda$ much lower than the Planck scale when a large number of particles $N$ are present. In four dimensions this bound comes with a precise scaling, namely that $\Lambda_{sp}\sim N^{-\frac{1}{2}}M_P$. One can derive this bound in several ways, here we follow an argument associated to black hole (BH) physics \cite{Dvali:2007hz}. The argument uses a theory with $N$ fields, each of which is charged under its own gauged discrete symmetry. The idea is then to form a BH with the maximal discrete charge, and to demand that it can consistently emit Hawking radiation. 

Let us run the argument at any number of
dimensions:\footnote{In the following the number of space-time dimensions in the uncompactified effective
theory is denoted by $d$, the number of compact dimensions are given by $d'$  and the number of total dimensions is called
$D=d+d'$.} we consider a theory of $N$ fields of mass $m$, each of them charged under a gauged $Z_2$ symmetry (so that the symmetry is at least $Z_2^N$). We can form a black hole charged under the full $Z_2^N$  group; its charge is measurable via Aharonov-Bohm experiments, so it must be radiated away at some point of the BH evolution. Thus, the BH must at least emit each of the $N$ quanta once. If $m$ or $N$ are very large, this has to happen when the BH is heavy, otherwise the emission is kinematically forbidden. However, when the BH is heavy, its Hawking temperature is small, so that the emission is unlikely. Thus, we must be able to ensure
\be
T_{BH} \gtrsim m \ , \quad M_{BH} > Nm \ . 
\ee
To be precise, at large $d$ the temperature $T_{BH}$ is not the relevant quantity to compare to $m$, because it is parametrically different from the frequency most radiated by the BH. The emission peaks at the energy $E_{BH}\sim\frac{d^2}{R_{BH}}$, while the temperature only scales as $T_{BH}\sim\frac{d}{R_{BH}}$ \cite{Emparan:2013moa}, so we must impose
\be
E_{BH} \gtrsim m \ , \quad M_{BH} > Nm \ . 
\label{conditionsForEvap}
\ee
The radius of the BH, in Planck units, is given in terms of the mass as follows
\be
R_{BH}\sim \sqrt{d}M_{BH}^{\frac{1}{d-3}} \ .
\ee 
Thus, 
\be
E_{BH}\sim d^{3/2}M_{BH}^{-\frac{1}{d-3}} \ .
\ee 
Using \eqref{conditionsForEvap}, at large $d$, we find $m \lesssim N^{-\frac{1}{d-2}}d^{3/2}M_P$. Given that all fields have mass $m$, a bound on the latter gives a higher bound on the cutoff,\footnote{We could also think of a tower of states of masses lying between $m$ and $Nm=\Lambda_{sp}$. Then, the conditions become $E_{BH} \gtrsim Nm$, $M_{BH} > N^2m$ and the species bound is unchanged.} so that we obtain the species bound at large (but fixed) $d$,
\be\label{speciesBound}
\frac{\Lambda_{sp}}{M_P}\sim N^{-\frac{1}{d-2}} \ .
\ee
This argument assumes that the charge must be Hawking-radiated away. It could also be that it is radiated at a stage of the BH evolution that is not anymore described by Hawking radiation. However, the species bound can also be derived by studying the perturbative running of $M_P$. By doing so, we find the same $N^{-\frac{1}{d-2}}$ scaling. Thus, at fixed but large $d$, the cutoff decreases very slowly when $N$ increases, and the species bound simply disappears at infinite $d$.

\subsection{The TCC and the distance conjecture at large D}

The second conjecture that we consider is the Trans-Planckian conjecture (TCC) \cite{Bedroya:2019snp}. It states that sub-Planckian quantum fluctuations should never expand beyond the Hubble horizon. Applied to scalar potentials $V$ in asymptotic field ranges, it yields
\be
\abs{V'}\geq \frac{2}{\sqrt{(d-1)(d-2)}}V \ .
\label{TCCforPotentials}
\ee
This condition becomes trivial at large $d$.

In addition, it has been conjectured in \cite{Andriot:2020lea} that the factor $\frac{1}{\sqrt{(d-1)(d-2)}}$ that appears in \eqref{TCCforPotentials} is a lower bound for the mass decay rate that appears in the swampland distance conjecture (SDC). More precisely, the swampland distance conjecture states that at large distances in field space, one expects towers of states that become light according to $m\sim e^{-\alpha \phi}$, with $\phi$ the field that undergoes a large displacement and $\alpha$ an unspecified, order one number. The claim in \cite{Andriot:2020lea} is that
\be
\alpha \geq \frac{1}{\sqrt{(d-1)(d-2)}} \ .
\label{lowerBoundLambda}
\ee
Although this bound has been conjectured based on compactifications of $10D$ supergravities, we can consider extending it to arbitrary dimensions. For instance, it holds for the tower of KK modes in the simple compactifications that we discuss later. At large $d$, it becomes unpredictive.

\subsection{The WGC at large D}

The weak gravity conjecture (WGC) can be expressed in arbitrary dimensions for arbitrary $p$-form gauge fields (with gauge coupling $e$) by reversing the extremality bound of charged black branes \cite{ArkaniHamed:2006dz,Heidenreich:2015nta}. The claim is that there should be a state of tension $T$ and quantized charge $Q$ such that
\be
\frac{p(d-p-2)}{d-2}T^2\leq e^2Q^2M_P^{d-2} \ .
\ee
Whatever $p$ is, this bound never becomes empty at large $d$. For a $1$-form field, it yields the usual
\be
m \lesssim eQM_P^\frac{d-2}{2} \ .
\label{WGClargeD}
\ee
The strongest bound comes when considering a $\frac{d}{2}$-form field:
\be
T \lesssim \frac{eQ}{\sqrt d}M_P^\frac{d-2}{2} \ .
\ee
When the charged brane is wrapped on the $\frac{d}{2}$ dimensions, it yields a particle which verifies
\be
m \lesssim \frac{eQ}{\sqrt d}M_P^\frac{\frac{d}{2}-2}{2} \ ,
\label{WGClargeD2}
\ee
which verifies again the WGC (note that it looks different than \eqref{WGClargeD}, due to the radion couplings in the compactified theory \cite{Heidenreich:2015nta}). Thus, the WGC is a first example of a conjecture that does not seem to weaken at large $d$, and that can even become very stringent. However, unlike the previous conjectures, there are unfixed parameters  in formulae like \eqref{WGClargeD2}, namely the charge and gauge coupling, so that it is unclear what the dependence with $d$ of the mass really is. For instance, the original version of the conjecture does not exclude a state whose charge scales with $\sqrt d$. Perturbative unitarity arguments may question the interpretation of that state as a particle, however it has been argued that the WGC state could be a BH, which can have a very large charge.


\section{A Large-D distance Conjecture}\label{largedconj}

We just saw that swampland conjectures having an unambiguous $d$-scaling seem to weaken at large number of dimensions. 
The interpretation of this fact is unclear, although we argued that it could be used to motivate an extension of the current set of swampland conjectures, 
or simply to reject the large-$D$ limit altogether in specific contexts. 
In order to gather more information in this direction, we focus on the swampland distance conjecture and then also on the AdS distance conjecture at large number of dimensions.

The  swampland  distance conjecture (SDC) \cite{Ooguri:2006in} states that  at large distances $\Delta$ in the field space 
of a $d$-dimensional (effective) quantum gravity theory there must be an infinite tower of states with mass scale $m$ such that 
\be
SDC:\quad m = M_P e^{-\Delta } \,.
\label{dsc}
\ee 
We assume that the spectrum of the tower is equidistant, which means that the number of states $N$ in the tower below the Planck mass is given as
\begin{equation}
N={M_P\over m}=e^\Delta\, .\label{number}
\end{equation}
As shown before in eq.(\ref{speciesBound}) for large number of dimensions, $N$ agrees with the number of states below the species scale.

The SDC predicts that EFTs which describe the bulk of moduli space break down at its boundaries.
This manifests itself by the appearance of towers of states whose masses are connected to large field displacements.  In quantum gravity or in 
string theory, the underlying reason why this happens is that objects revealing the higher-dimensional nature of the theory, or its 
stringy nature, become part of the low-energy modes, as happens for instance in  a decompactification limit or a tensionless limit  \cite{Lee:2019wij}. 

Decompactification in string theory must be understood in a broader sense than the usual field theory one. Indeed, while a large volume limit is well described in the field theory limit of string theory (which already captures the presence of KK modes necessary to "build" the compact geometry), a small volume limit often displays the stringy nature of the theory. Indeed, winding modes become light and take over the KK modes, which become heavy, to reconstruct a (T-)dual geometry. 

Therefore, concerning the emergence of winding modes
we would like to insist on the following: we could have either noticed that the small volume limit corresponds to a large distance in field space in the EFT and used the SDC, or we could have simply remarked that the KK modes become heavy, which weakens our geometric picture. Then, we would have anticipated that there is a dual formulation where a geometrical picture emerges again, or that the small volume limit is not consistent in the first place. 
 Indeed, an immediate consequence of the SDC is that,
if there happens to be an infinite distance limit in field space without a corresponding tower of light states satisfying \eqref{dsc}, then it cannot be possible to approach this infinite distance point. 

Therefore, let us revert the logic of the SDC and define, for a given tower tower of states $\left|i\right>$, the quantity $\Delta_i$ (the ``distance'') as the negative logarithm of its typical mass scale $m_i$, i.e.
$\Delta_i \sim - \log  m_i$.
Following \eqref{dsc} we normalize $\Delta_i$ such that $m_i = M_{P}$ for $\Delta_i = 0$.
At some infinite distance point in field space, $\Delta_i$ agrees with the distance computed from the $\sigma$-model metric.
Analogously, if there is a dual tower of states $\left|\tilde \imath\right>$ we define the dual distance $\widetilde \Delta_i \sim - \log \widetilde m_i$.
Following our definition, $\Delta_i$ and $\widetilde \Delta_i$ are not the same (even though typically closely related).
This is consistent with the SDC as the two towers become light at different points in moduli space.
For example, for string theory on a circle with $\left|i\right>$ and $\left|\tilde \imath\right>$ given by Kaluza-Klein and winding modes we have $\Delta_i \sim \log R$ and $\widetilde \Delta_i \sim - \log R$.
Similarly for compactification on a 2-torus, $\left|i\right>$ and $\left|\tilde \imath\right>$ given by two different Kaluza-Klein modes, and $\Delta_i \sim \log (\Im \tau)$ and $\widetilde \Delta_i \sim - \log (\Im \tau)$,
where $\tau$ is the complex structure modulus of $T^2$.

On the basis of the SDC and following these observations, we now demand that,  for a particular tower $\left|i\right>$,  an "infinite distance" limit $\Delta_i\rightarrow -\infty$ is obstructed, i.e.~not allowed,
unless there is a dual tower $\left|\tilde \imath\right>$, which becomes light in this limit. 
It would be tempting to dub this requirement the \emph{Negative Distance Conjecture (NDC)}.
Note that in general there is not necessarily a simple relation between $\Delta_i$ and the geometric distance in scalar field space, so we might not directly be able to use the SDC to infer anything about the limit $\Delta_i \rightarrow - \infty$. In the case of KK modes, the limit $\Delta_{KK} (\phi)\rightarrow -\infty$ implies $\phi \rightarrow -\infty$ and the SDC says that there must be a second tower of states or $\phi \rightarrow -\infty$ cannot be possible. With the NDC we extrapolate from the KK modes behaviour a general pattern assumed to be true for any tower of states. 

Let us now go back to our discussion of large-$D$ EFTs. In general, the quantity $\Delta$ that determines the mass scale of the tower in \eqref{dsc} is a field dependent quantity, $\Delta=\Delta(\phi)$, where $\phi$ denotes the scalar fields in the EFT.
Typically they 
correspond to geometric length
parameters like the size of a compact space.
However, the mass scale of the tower can also depend on the number of space-time dimensions in a non-trivial way.
As we will extensively discuss later, this is the case for the KK spectrum.
Therefore, we introduce in addition to  $\phi$ the  space-time dimension $D$ 
as a new swampland variable that determines the spectrum and the validity regime of the EFT.

In particular, we focus our discussion on the dependence of the masses of known towers of states on the dimension of space-time. Said differently, we study the distance $\Delta$ as a function of the number $D$ of space-time dimensions, i.e. $\Delta=\Delta(\phi,D)$.
As a consequence, infinite distances have to be extended to incorporate the $D$-direction. In particular, even for single field models, there are now different ways to go to infinite distance, since $D$ can be taken to be fixed or to vary. The usual towers of states, such as KK modes, may not have the expected behavior once $D$ varies, even when the distance restricted to the field space of $\phi$ becomes infinite.

What does this extension of the parameter space lead to? We do not have a dynamical theory of $D$, i.e.~there is no  kinetic term for $D$ in an EFT, and we cannot use the SDC 
in the standard way.
But on the basis of the previous discussion, namely utilizing the NDC,
we are now ready to propose the following \emph{Large Dimension Conjecture (LDC)}:%

\vskip0.3cm
\noindent\emph{\noindent
If there is a tower of states $\left |i\right>$, the corresponding ``distance'' $\Delta_i(\phi,D)$ must be a positive  function of the EFT fields $\phi$ and the dimension $D$,
\begin{equation}
LDC:\quad\Delta_i(\phi,D)\geq0 \,,\label{condition}
\end{equation}
unless there is a dual tower of states $\left |\tilde \imath\right>$ such that the dual distance $\widetilde\Delta_i(\phi, D)$ becomes positive when $\Delta_i(\phi,D)$ changes its sign and takes negative values.
}

In simple cases one is dealing with a two-parameter space of backgrounds, 
the number $D$ of space-time dimensions and one geometric size parameter of the background, denoted by  $\phi \sim \log R$. 
In this case we can give a simple interpretation of the LDC:
the geometric interpretation of space-time breaks down when $\Delta_{KK} (R,D)\rightarrow -\infty$.
Therefore, we conclude that something must happen in this limit.
Either we expect a dual geometry with a dual tower of winding modes to emerge or the limit is not consistent.

In the two parameter case we can look for one-dimensional directions of constant $\Delta$.
These lines will  separate the parameter space into a region in which the tower mass scale is smaller than one, i.e.~positive $\Delta$, and another region in which
the tower mass scale is larger than one and $\Delta$ is negative.
Forbidding this region then provides a possible bound on a combination of $D$ and $R$.

More precisely, for fixed $D$, the LDC provides a $D$-dependent bound on $R$, which in general is stronger than just requiring 
$R>1$.
But in the context of large-$D$ gravity, we also like to treat $D$ as a free variable and then the LDC provides,  for fixed $R$, a  bound  on the number of space-time dimensions, namely a critical dimension $D_0$.
In fact, treating $D$ as a distance variable and keeping $R$ fixed, $\Delta(D)$ should measure the distance 
 between backgrounds of different dimension.\footnote{One could try to derive $\Delta(D)$ as the distance between space-time
backgrounds using the 
geometric distance formula
$\Delta \sim\int  \left( \frac{1}{{\cal V}_M} \int_M \sqrt{g} \tr[\left(g^{-1}\frac{\pa g}{\pa \tau}\right)^2] \right)^{\frac12}\text d\tau
$,
where ${\cal V}_M$ denotes the volume of space-time.
The  parameter $\tau$ is given by the number of space-time dimensions $D$, i.e.~one 
should compute the distance for metric variations with respect to $D$.
Alternatively, as in \cite{Kehagias:2019akr},
 one can also try to set up geometric flow equations, like the Ricci flow, where one considers the flow of the background with respect to $D$. This will be discussed elsewhere  \cite{workinprogress}.}
As we will see for the KK tower or for black holes the LDC means
that $D_0$ is determined by the special dimension in which the space-time volume becomes basically smaller than one in Planck units, which should not happen in quantum gravity,
unless there is a dual description of the theory.

Let us see what the LDC means for the canonical volume modulus of compactification spaces with a large number of dimensions.
We consider a $D$-dimensional space-time, which splits into a non-compact space $M_{d}$ and a compact space $K_{d'}$ of dimensions $d$ and $d'$ respectively, like for $AdS_{d} \times S^{d'}$, so that $D=d+d'$.  
The ansatz for the $D$-dimensional metric reads
\begin{equation}
\d s^2_D = e^{2 \beta \varphi} d s^2_{M_{d}} + e^{2 \phi} d s^2_{K_{d'}} \,,
\label{ansatzdTimesd'}
\end{equation}
where we split the volume modulus $\phi(x^\mu) = \phi_0 + \varphi(x^\mu)$ into a constant piece $\phi_0$ and its dynamical part $\varphi$ which 
depends only on the coordinates $x^{\mu}$ of the ``external'' space, but not on the coordinates $y^m$ of the ``internal'' $K_{d'}$.
In addition, we normalize its background value $\phi_0$ and the volume of  $K_{d'}$ according to
\begin{equation}
{\cal V}_{d'}\equiv \left(M_P^{(D)}\right)^{d'}\int_{K_{d'}} d^{d'} y \sqrt{\det(e^{2\phi_0}g_{K_{d'}})} =e^{d'\phi_0} \, .
\label{referenceRadiusNormalization}
\end{equation}
Consequently, ${\cal V}_{d'}$ is dimensionless and measured in units of the $D$-dimensional Planck mass.

After reduction of the $D$-dimensional Einstein-Hilbert term, we reach the lower-dimensional Einstein frame  by fixing
\beq
\beta = - \frac{d'}{d-2} \,.
\eeq
The relevant terms in the $d$-dimensional effective action look like
\begin{equation}
\mathcal{L} = \frac{\bigl(M_P^{(d)}\bigr)^{{d}-2}}{2} \int d^{d} x\sqrt{-g_{M_d}} \left[\mathcal{ R}_{d} - \frac{d'(D-2)}{d-2} \left|\partial \phi \right|^2  \right] \,.
\end{equation}
The lower dimensional Planck mass  is related to the higher dimensional Planck as
\be
M_{P}^{(d)}=M_{P}^{(D)} {\cal V}_{d'}^{1\over d-2} \ .
\label{invariantMP}
\ee
Let us finally introduce the canonically normalized field
\begin{equation}
\Phi=  \sqrt {\frac{d'(D-2)}{d-2}}   \phi=\sqrt {\frac{D-2}{d'(d-2)}} \log{\cal V}_{d'}
\,.
\label{canonicalFieldGeneral}
\end{equation}

According to the infinite distance conjecture, the normalized field $\Phi$ is related to a tower of states with masses $m$ given by
\begin{equation}
m\sim M^{(d)}_P e^{-\alpha(d,d') \left|\Phi\right|}\, ,
\end{equation}
Here $\alpha(d,d')$ is a dimension dependent constant, which we will determine in the following.
For the infinite distance point $\Phi \rightarrow \infty$ the relevant mass scale $m$ is just given in terms of the KK mass scale,
\beq\begin{aligned}
m_{KK}\approx {\cal V}^{-{1\over d'}}_{d'}M_P^{(D)} \approx M^{(d)}_P{\cal V}_{d'}^{-{D-2\over d'(d-2)}} 
= M^{(d)}_P \exp\left\lbrack-\sqrt{\frac{D-2}{d'(d-2)}}\Phi\right\rbrack\, .
\label{KKmasses}
\end{aligned}\eeq
It follows that the constant $\alpha(d,d')$ is given by
\begin{equation}
\alpha(d,d')=\sqrt{\frac{D-2}{d'(d-2)}} \,.
\end{equation}
Consistently with the infinite distance conjecture, the KK tower becomes light in the limit of large $\Phi$. Given our definition of $\Phi$, this limit can be reached in the usual way, namely as a large-$\Phi$ limit.
On the other hand we point out that, for fixed $\Phi$, the KK masses approach $ M_P^{(d)}$ at $d,d' \rightarrow \infty$.

In any case, using our definition \eqref{dsc}, the distance $\Delta$ and $\Phi$ are related as $\Delta=\alpha\Phi$, and 
the LDC requires that $\Delta(d,d')>0$, i.e.
$m_{KK}<M_P^{(d)}$, which means
\begin{equation}
\Phi>0\, ,
\label{generalConditionCompactification}
\end{equation}
unless one identifies a dual tower of states that becomes light in the limit under consideration.


\section{The LDC for AdS at large D}\label{sectads}

\subsection{The AdS distance conjecture and KK spectrum in arbitrary number of dimensions}

Now let us briefly recall the  anti-De Sitter distance conjecture (ADC) \cite{Lust:2019zwm}, which is obtained by comparing AdS spaces with different cosmological constants $\Lambda$, i.e. varying the metric with respect to $\Lambda$.
The ADC is a bit different from the SDC in the sense that it deals with the distance between not continuously connected backgrounds, 
   which implies that the scalar fields are not anymore massless scalars but possess a potential. Different discrete vacua are labeled by different values of the cosmological constant or
   the AdS radius, which are related to different, discrete
   flux quantum numbers.

Then ADC states that the limit of a small AdS cosmological constant, $\Lambda\rightarrow0$,  is at infinite distance in the space of AdS metrics, and that it is related to an infinite tower of states with typical masses that behave as
\begin{equation}
{\rm ADC}\colon\quad m_{AdS}\sim \left|\Lambda\right|^a\, ,\label{adstower}
\end{equation}
with $a={\cal O}(1)$.  The strong version of the ADC proposes that for supersymmetric backgrounds  $a=1/2$.
The corresponding distance is given in terms of the logarithm of $\Lambda$,
\begin{equation}
\Delta_{AdS}=-a\log \left|\Lambda\right|\, .
\end{equation}
Via \eqref{dsc}, the distance $\Delta_{AdS}$ immediately leads to the tower behaviour (\ref{adstower});  as long as $a$ is positive,  the distance $\Delta_{AdS}$ is positive
 in the limit of small $\Lambda$ and the mass scale $m_{AdS}$ is decreasing in this limit.
 So in the limit of vanishing cosmological constant, i.e.~large AdS radius, AdS quantum gravity as an effective field theory can only co-exist together with an infinite tower of additional  massless states.
 This  implies that pure AdS quantum gravity is in the swampland.
 The tower of states is typically provided by the KK modes of an extra compact $d'$-dimensional space, e.g. with backgrounds of the form $AdS_d\times S^{d'}$. 
 For  the case of the strong ADC with  $a=1/2$ it follows that there is no scale separation between the mass scales of $AdS_d$ and  $S^{d'}$.

In the following we will consider  AdS spaces in arbitrary dimensions $d$:
\begin{equation}
\D s^2 = R_A^2 \left( - \left(\cosh \rho\right)^2\D t^2 + \D\rho^2 + \left(\sinh \rho\right)^2 \D \Omega_{d-2}^2 \right) \;.\label{ads2}
\end{equation}
Here $R_A$ is the AdS radius which is related to the cosmological constant $\Lambda_{AdS_d}$ as
\begin{equation}
\Lambda_{AdS_d}=-{(d-1)(d-2)\over 2R_A^2}\, .\label{lr0}
\end{equation}

Since AdS$_d$ alone is not a viable background, we consider a background space of the form
\begin{equation}
{\cal M}_D=AdS_d\times S^{d'}\,\, {\rm with}\,\,\ d+d'=D.\label{freundrubin}
\end{equation}
Recall that the Ricci scalars of AdS$_d$ and of the compact $d'$-dimensional sphere $S^{d'}$ of radius $R_S$ are given  by 
\begin{equation}\label{ras}
{\cal R}_A=-{d(d-1)\over R_A^2}\, ,\quad
{\cal R}_S={d'(d'-1)\over R_S^2}\, .
\end{equation}
This background is obtained from the so-called Freund-Rubin compactifications from $D$ dimensions, which is a solution of the $D$-dimensional Einstein equations without a cosmological constant by turning on a non-vanishing $d$-form gauge field strength in the AdS-space:\footnote{By Hodge duality, it can equally lie on the $S^{d'}$ factor, and read
\begin{equation}
F^{\mu_1\dots\mu_{d'}}= \frac{\epsilon^{\mu_1\dots\mu_{d'}}}{\sqrt{g_S}}f\, .
\end{equation}}
\begin{equation}
F^{\mu_1\dots\mu_d}= \frac{\epsilon^{\mu_1\dots\mu_d}}{\sqrt{-g_A}}f\, ,
\end{equation}
such that the Ricci scalars are given by
\be
{\cal R}_A=-\frac{d(d'-1)}{D-2}f^2 \ , \quad {\cal R}_S=\frac{d'(d-1)}{D-2} f^2 \ ,
\ee
in units of $M_P^{(D)}$, the $D$-dimensional Planck mass. These expressions correlate the radii of the two factors:
\be
(d-1)R_S=(d'-1)R_A \ .
\label{radiiRelationFR}
\ee

In the next step we can compute the spectrum of KK states in the effective AdS$_d$ theory which originate from the compactification on  $S^{d'}$.
This spectrum obeys the following equation (note that from now on, we measure the radii in units of D-dimensional Planck
mass $M_P^{(D)}$)
 \begin{equation}
m_{l,KK}^2=\Bigl(M_P^{(D)}\Bigr)^2~{l(l+d'-1)\over R_S^2}\, ,
\end{equation}
where the integer $l$ labels the KK momentum.
Comparing this with the AdS cosmological constant $\Lambda$, the associated 
 KK mass scale ($l=1$) is then given by
\begin{equation}\label{mk}
m_{KK}^2=-2{(d-1)d'\over (d-2)(d'-1)^2}\Lambda_{AdS_d} \ ,
\end{equation}
so we see that $\alpha=1/2$ and the strong ADC still holds in arbitrary number of  dimensions.

\subsection{The large D behaviour}

With the LDC, one can now ask what happens when the dimension $d$ of AdS or/and $d'$ of $S$ is varied and gets large.
As we will see, certain limits for $d,d'$ lead to violations of the LDC if we assume that the tower predicted by the ADC is again the tower of KK modes of a Freund-Rubin compactification. From that, we conjecture either a bound relating the dimension of spacetime and geometrical quantities such as radii, or we predict the presence of a dual tower that should become light in the same limit.

A relevant quantity  for the following discussion is the volume of the $d'$-dimensional unit sphere, given by
\begin{equation}
\Omega_{d'}={2\pi^{(d'+1)/2}\over \Gamma\bigl({d'+1\over 2}\bigr)}\, .\label{volumeunitsphere}
\end{equation}
For large $d'$ this behaves as
\begin{equation}
\Omega_{d'}\sim \sqrt 2 e\Biggl(\sqrt{{2\pi e\over d'}}\Biggr)^{d'}\rightarrow 0\, ,\label{unitsphere}
\end{equation}
and it therefore becomes increasingly small for large $d'$, i.e. $\Omega_{d'}^{1/d'}\rightarrow \frac{1}{\sqrt{d'}}$ for $d'\rightarrow\infty$.
In fact, the area of the unit sphere becomes of order one for 
\begin{equation}
 d'\sim2\pi e\sim  17\, .
 \end{equation}
 Including also the radial dependence, the volume of the $d'$-dimensional sphere of radius $R_S$ for large $d'$  is given by ${\cal V}_{d'}=R_S^{d'}\Omega_{d'}$.

\subsubsection{Small number of internal dimensions}

Let us first consider the case where the number of dimensions $d'$ of the sphere is fixed\footnote{As a side remark, it is also interesting to consider the case where the AdS background is fully fixed, namely the case of fixed $d$, fixed cosmological constant $\Lambda_{AdS_d}$ (or AdS radius $R_A$) and fixed $d$-dimensional Planck mass. We then take the large $D$, i.e.~large $d'$, limit (that implies that the $D$-dimensional Planck mass vanishes asymptotically). In this limit, it is easy to see that the KK mass scales as
\begin{equation*}
m_{KK}^2\sim \frac{\Lambda_{AdS_d}}{D} \ .
\end{equation*}
Thus, a tower of states enters the AdS EFT simply because we considered the large $D$ limit. This is due to the fact that the canonical $\Phi$ in \eqref{canonicalFieldGeneral} has to vary for the AdS background to remain fixed, and this behaviour is captured by the SDC. In this limit, the KK modes are light and the LDC does not bring new information.} and where we take only  $d\sim D\to\infty$. The $d$-dimensional Planck mass reads
\beq
M_P^{(d)}=M_P^{(D)}{{\cal V}_{d'}}^{\frac{1}{d-2}}\sim M_P^{(D)} \ .
\eeq
Consequently, the KK masses are given by their expression in units of $M_P^{(D)}$, given in eq.(\ref{mk}),
\beq
m^2_{KK}\sim -{2\over (d'-1)}\Lambda_{AdS_d} \sim-\Lambda_{AdS_d}\,.
\eeq
The corresponding AdS distance is therefore
\begin{equation}
\Delta_{AdS}=-{1\over 2}\log \Lambda\, .
\end{equation}
In terms of $R_S$ these relations look like
\beq
m^2_{KK} \sim\Bigl(M_P^{(D)}\Bigr)^2~{1\over R_S^2}\, ,\quad \Delta_{AdS}=\log R_S\,.
\label{mKKandRslargedfixedd'}
\eeq
We see that $m_{KK}$ and $\Delta_{AdS}$ do not depend on the number of space-time dimensions. Actually this is the same situation as for the KK spectrum arising
from the compactification on a single circle.

On the other hand, in terms of $R_A$, the analogous relations look like 
\beq
m^2_{KK} \sim\Bigl(M_P^{(D)}\Bigr)^2~{D^2\over R_A^2}\, ,\quad \Delta_{AdS}=\log \biggl({R_A\over D}\biggr)\,.
\eeq
Therefore the KK mass scale becomes heavier than the Planck mass for $D>R_A$. Using \eqref{radiiRelationFR}, the latter condition is equivalent to $R_S<1$, consistently with \eqref{mKKandRslargedfixedd'}.

\subsubsection{Large number of internal dimensions}

Next let us assume that the radii of the two subspaces are  of the same order, i.e.
$R_S=R_A=R$. It then follows that also the respective numbers of dimensions must be the same: $d=d'=D/2$.
It also follows that in terms of the flux number $f$ we have that $R\simeq D f^2$.
In the limit of large dimensions, using eq.(\ref{mk}), the KK mass scale 
behaves as
\begin{equation}\label{mads}
m_{KK}^2\sim{|\Lambda_{AdS_d}|\over D}\,.
\end{equation}
Instead of expressing the KK scale  in terms of $\Lambda_{AdS_d}$, we can also express it in terms of the 
radius $R$:
\begin{equation}
m_{KK}^2
\sim \Bigl(M_P^{(D)}\Bigr)^2~\biggl({1\over {\cal V}_{d'}}\biggr)^{2/d'}
\sim\Bigl(M_P^{(D)}\Bigr)^2~ {D\over R^2}\,,\label{KKMasslargeD}
\end{equation}
In this relation $m_{KK}$ is measured in units of the higher dimensional Planck mass $M_P^{(D)}$. 
In our case, the Planck mass in the effective AdS-theory behaves as
\begin{equation}
M_P^{(d)}\sim M_P^{(D)}\biggl({R\over\sqrt D}\biggr)\, .
\end{equation}
Therefore in terms of $M_P^{(d)}$, the KK masses become
\begin{equation}
m_{KK}^2\sim\Bigl(M_P^{(d)}\Bigr)^2~ \biggl({D\over R^2}\biggr)^2\, ,
\label{KKMasslargeD2}
\end{equation}
where $R$ is measured in units of $M_P^{(D)}$.
Comparing with the $D/2$-dimensional effective action, the normalized field $\Phi$ depends on $R$ and $D$ as
\begin{equation}
\Phi(R,D)\simeq \sqrt D\log\biggl({R\over \sqrt D}\biggr) \,,
\end{equation}
and becomes therefore negative for $D > R^2$.

So we see  that, at fixed radius, all KK masses become super Planckian above a certain critical dimension.
The corresponding AdS distance is given as
\beq
 \Delta_{AdS}=2\log \biggl({R\over \sqrt D}\biggr)\,,\label{AdSdistanceLargeD}
\eeq
and it is positive for $D$ only smaller than a certain critical dimension $D_0$, namely
\begin{equation}
D\leq D_0=R^2\, ,\label{upperD}
\end{equation}
where all KK tower starts below the Planck scale. This condition is nothing but the application of  \eqref{generalConditionCompactification} to the case of $AdS_{D/2}\times S^{D/2}$. It gives us a non-trivial constraint on $R$ and $D$ because the volume of the sphere has a non-trivial dimension dependence.

On the other hand, if this condition is violated, all KK masses are super Planckian and 
then the notion of space-time is in general lost.
Actually the KK modes are part of the space-time geometry - they build geometry -  and certain KK towers
must start below the Planck mass, if one wants to have a geometrical interpretation of space-time. 
Therefore,  a  KK tower that starts above the Planck mass appears pathological in quantum gravity and one can conjecture that this regime of the parameter space is in the
swampland (unless there is a dual tower). This observation is the reasoning for the LDC, which leads to the  upper bound (\ref{upperD}) on $D$ for $AdS_d\times S^{d'}$ geometries. 

However note that for AdS space-times of radius $R$ (see \eqref{ads2} for the metric) many of the curvature invariants do not vanish, for instance:
\beq
\bal
{\cal R}_{\mu\nu\sigma\rho}{\cal R}^{\mu\nu\sigma\rho}= \frac{2D(D-1)}{R^4} \ , \quad {\cal R}_{\mu\nu}{\cal R}^{\mu\nu} &= \frac{D(D-1)^2}{R^4} \ ,\\
{\cal R}^2=\frac{D^2(D-1)^2}{R^4} \ .&
\eal
\eeq
So demanding that the curvature stays less than the Planck mass, i.e. ${\cal R}^2\lesssim M_P^4$,  implies $R\gtrsim D$.

\vskip0.2cm

In summary, for the case with both $AdS_d$ and $S^{d'}$ having large dimensions there are three regions,\footnote{For the case with only AdS$_d$ having a large dimension, but the sphere having a finite dimension there are two regimes:

$D<R:$  small curvature and good notion of space-time geometry.

$R< D$: large curvature and no good notion of space-time geometry, the  KK tower is above the Planck scale.} when comparing $R$ and  $D$:

A: $D<R$:  small curvature, light KK tower and good notion of space-time geometry.

B: $R < D < R^2$: large curvature but still good notion of space-time geometry, i.e. the KK tower is still below the Planck mass.

C: $R^2 <D$: large curvature and no notion of space-time geometry, as the  KK tower is above the Planck mass.

Given that the LDC only brings something non-trivial in regimes of large curvature, we cannot exclude that our bounds are too naive and are affected by strong gravitational corrections. It would be interesting to investigate the constraints imposed by the LDC in geometries where the curvature is controllable, to firmly establish the bounds on $R$ and $D$, the typical size of the geometry and its dimension \cite{workinprogress}.

We would like to conclude with a remark about IIB string theory on $AdS_5 \times S^5$ or M-theory on $AdS_4 \times S^7$ or $AdS_7 \times S^4$ backgrounds. 
There are no winding modes on the sphere, however there exists the classical limit of small sphere radius where the KK modes decouple. There,  $\Delta_{KK}<0$ without any dual tower, which is excluded by the LDC, and indeed flux quantization on the sphere forbids that we take the small radius limit or the large dimension limit, respectively.
The flux quantization condition emerges from the existence of objects with Planck-scale charge, namely $D3$, $M2$ or $M5$ branes.
Equivalently, these backgrounds can also be obtained as the near-horizon geometry of these branes, whose number cannot be smaller than one.
It is thus conceivable that the LDC is connected to the existence of such objects in large $D$, as also required by the completeness conjecture \cite{Polchinski:2003bq}.

\section{The black hole entropy distance at large $D$}\label{sectbh}

As it was recently shown in \cite{Bonnefoy:2019nzv}, the limit of large entropy, ${\cal S}\rightarrow\infty$, is at infinite distance in the space of  black hole metrics. Therefore, similarly to the ADC, 
it was  argued \cite{Bonnefoy:2019nzv} that there exists a black hole entropy distance conjecture (BHEDC) such that
\begin{equation}\Delta_{bh}=\alpha\log{\cal S}\, ,
\end{equation}
stating that the large entropy limit of black holes is also followed by a tower of light modes:
\begin{equation}
m_{bh}\sim{1\over{\cal S}^\alpha}\, .
\end{equation}
A possible interpretation of this tower of states is closely related to the black hole micro states.

For black holes at large space-time dimensions we can use the result for 
 the black hole entropy at large $D$,
which is given by the following expression \cite{Emparan:2013moa}:
\begin{equation}\label{largeDentr}
{\cal S}\sim\biggl({R_h\over \sqrt{D}}\Biggr)^{D-2}\, .
\end{equation}
Here $R_h$ (in Planck units) plays the role of the radius of the black hole horizon.
For $D\rightarrow \infty$ the entropy becomes very small. Note that the $1/\sqrt D$ in the denominator again originates from  the area of the unit sphere in large dimensions in the same way as for the KK mass scale
in the previous section.
So comparing the large-$D$ entropy with the distance $\Delta_{bh}$ we obtain
\be
\Delta_{bh} = \alpha\log {\cal S}=\alpha(D-2)\log  \biggl({R_h\over \sqrt{D}}\Biggr)        \,. 
\label{dismet3}
\ee
We see that the black hole distance has the property that it is only positive in a finite interval of $D$,
\begin{equation}
\Delta (D)_{bh}\geq 0\quad{\rm for} \quad 2\leq D\leq R_h^2\, ,
\end{equation}
with a maximum of the distance somewhere between $D=2$ and $D=R_h^2$.
Therefore the LDC again implies
 the following  critical  dimension:
 \begin{equation}
 D_{0}\sim R_h^2\, .
\end{equation}
Respecting the bound means that  the black hole entropy should be larger than one.
Now the LDC is based on the  argument that black holes with entropy smaller than one
are not consistent objects in quantum gravity.
Note that the case $D=2$ is special: the distance is zero and the black hole tower contains only one state below the Planck mass.

\section{Discussion}

In this paper we have considered some swampland aspects for gravity theory in an arbitrary number and in particular in a large number of space-time dimensions.
The discussion was mainly focussed on the large distance conjecture with KK modes, whose masses exhibit a non-trivial dependence on the number of space-time dimensions.
In addition we also looked at the black hole distance at large $D$.
Based on these results
we
formulated a new large-$D$  conjecture stating that certain distance functionals of the swampland distance conjecture must be positive as a function of the
number of space-time dimensions. Being certainly satisfied in critical string theory, the LDC is supposed to provide a general swampland condition for quantum gravity in arbitrary dimensions.
Applying it to AdS spaces and to black hole geometries, it then leads to an upper critical dimensions from the requirement that the volume of the compact space must be larger than one or, in the context of black holes, that the black hole entropy should be larger than one at large D.
Since $D$ is a discrete number, the existence of a critical dimension would strictly spoken also require that the AdS sizes and the black hole entropies that saturate the LDC bound are also quantized.
In fact the possibility of discrete and quantized black hole entropies was considered in the context of quantum black holes in \cite{Bekenstein:1995ju,Mukhanov:2018tzo}.

As we have seen from the masses of the KK modes the critical dimension depends on the size of the compact space, e.g.
$D_0=R^2$ in the case of large dimensions for the internal and AdS spaces. 
In particular, for 
 large AdS spaces (or black holes), i.e. $R\gg 1$ (or $R_h\gg1$), the critical dimension $D_0$ is very large. 
 But what about  the possible existence of a critical dimension, which is independent from the length scale $R$?
For that let us impose  the much stronger requirement that all possible backgrounds such as 
all possible AdS spaces 
 or all possible black holes 
must satisfy the LDC. Clearly, the strongest bound arises
for
Planck size compact spaces $R\sim{\cal O}(1)$ or Planck size black holes $R_h\sim{\cal O}(1)$. Then one gets an
absolute bound on $D_0$ of the order ${\cal O}(2-20)$.
Another speculative way to obtain a critical dimension is via 
 the exact formula of volume of the unit sphere $S^{D/2}$ in eq.(\ref{volumeunitsphere}):
$\Omega_{D/2}$ is an increasing  function  with $D$ for small $D$ and then decreases for large $D$ with a maximum volume at $D/2=6$.
So the $AdS_{5}\times S^{5}$ background with $D/2=5$ in superstring theory is tantalizingly close to  the unit sphere of maximal volume.

We also like to recall
that for the case of the curved $AdS_d\times S^{d'}$ backgrounds
 the KK masses in AdS space start to become heavier than the Planck mass in the regime where the space-time curvature is already large.
 One might therefore wonder if our "naive" estimates for the KK masses are still valid, or if there are non-negligible gravitational contributions to the KK masses that keep them always below
 the Planck mass. 
 On the other hand for the flat background, there are no curvature corrections and hence the KK spectrum at large dimensions would be under control.
 So it would be very interesting to see how the large $D$ spectrum for flat backgrounds depends on the number of space-time dimensions, like for tori or Ricci flat spaces at high number of dimensions,
 or also for other high-dimensional spaces with non-vanishing torsion classes \cite{workinprogress}.

As discussed before, the distance conjecture often contains the expectation that at large distances in field space one ends up with a dual description, where states  getting light at large distances are accompanied by a dual tower of heavy states for large field values.
Moreover often there is a dual limit in field space, with a dual distance $\tilde\Delta$, where the situation is reversed. 
Here
 we have seen that states get heavy at large $D$. Therefore, non-perturbative branes wrapped around certain high-dimensional cycles might become light in the large-$D$ limit \cite{workinprogress}.
In view of the possible presence of these non-perturbative states, it would be interesting to investigate the existence of a novel non-perturbative duality symmetry in quantum gravity,
denoted by {\sl D-duality}, which could act in  following way on $R$ and $D$: 
\begin{equation}
(\sqrt D/R) \,  \longleftrightarrow \, (R/\sqrt D)\, .
\end{equation}
One may then consider two limiting cases with either the dimension or the size fixed. The former could be realized by known dualities: for superstring theory in the critical dimension $D_0=10$, D-duality would act on the internal radius, e.g. in the four-dimensional heterotic strings on $T^6$ 
it may act
like S-duality \cite{Font:1990gx} exchanging strings with wrapped NS-5 branes. The latter much more intriguing but also exotic case
would exchange quantum gravity theories in  different dimensions.

Finally, how the LDC is precisely related to the critical dimension in string theory is not obvious at the moment.
One possible way to relate the swampland bounds on the number of space-time dimensions to the critical dimension in string theory might be  via recent arguments about entropy bounds and
the unitarity of scattering amplitudes \cite{Dvali:2020wqi}.

\vspace{10px}
{\bf Acknowledgements}
\vskip0.1cm
\noindent

We thank Gia Dvali, Eran Palti, Cumrun Vafa and Timo Weigand
for very useful discussions.
Q.B. is supported by the Deutsche Forschungsgemeinschaft under Germany's Excellence Strategy  EXC 2121 ``Quantum Universe" - 390833306.
L.C. is supported by the ERC Advanced Grant \textsl{High-Spin-Grav}.
The work of D.L. is supported  by the Origins Excellence Cluster.
The work of S.L.~is supported by the ANR grant Black-dS-String ANR-16-CE31-0004-01, the ERC grant 772408 ``String landscape'' and the US National Science Foundation grant PHY-1915071.


\end{document}